\documentclass{aa}  

\usepackage{graphicx}
\usepackage{txfonts}
\usepackage{lineno,hyperref}
\usepackage{amssymb}
\usepackage{xcolor,colortbl}
\usepackage{graphicx}

\usepackage{array}
\usepackage{lscape}
\usepackage{rotating}
\usepackage{longtable}
\usepackage{multirow}
\usepackage{color}
\usepackage{amsfonts}
\def\fm{$f_{\rm m}$}
\def\kf0{$kf_{0}$}

\begin{document}

   \title{The SERMON project: 48 new field Blazhko stars and an investigation of modulation-period distribution}


   \author{M. Skarka,
          \inst{1,2,3}
          J. Li\v{s}ka\inst{2,3},
          R.\,F. Auer\inst{3}, Z. Prudil\inst{2}, A. Jur\'{a}\v{n}ov\'{a}\inst{2},
          \and
          \'{A}. S\'{o}dor\inst{1}          
          }

   \institute{Konkoly Observatory, MTA Research Centre for Astronomy and Earth Sciences, Konkoly Thege Mikl\'{o}s \'{u}t 15-17, H--1121 Budapest, Hungary; \email{marek.skarka@csfk.mta.hu}
         \and
             Department of Theoretical Physics and Astrophysics, Masaryk University, Kotl\'{a}\v{r}sk\'{a} 2, 611 37 Brno, Czech Republic
         \and
             Variable Star and Exoplanet Section of the Czech Astronomical Society, Vala\v{s}sk\'{e} Mezi\v{r}\'{i}\v{c}\'{i}, Vset\'{i}nsk\'{a} 78, Vala\v{s}sk\'{e} Mezi\v{r}\'{i}\v{c}\'{i}, 757~01, Czech Republic\\
             }

   \date{Received 6 May 2016 / Accepted 17 June 2016}

 
  \abstract
  {}
   {We investigated 1234 fundamental mode RR Lyrae stars observed by the All Sky Automated Survey (ASAS) to identify the Blazhko (BL) effect. A sample of 1547 BL stars from the literature was collected to compare the modulation-period distribution with stars newly identified in our sample.}
   {A classical frequency spectra analysis was performed using \textsc{Period04} software. Data points from each star from the ASAS database were analysed individually to avoid confusion with artificial peaks and aliases. Statistical methods were used in the investigation of the modulation-period distribution.}
   {Altogether we identified 87 BL stars (48 new detections), 7 candidate stars, and 22 stars showing long-term period variations. The distribution of modulation periods of newly identified BL stars corresponds well to the distribution of modulation periods of stars located in the Galactic field, Galactic bulge, Large Magellanic Cloud, and globular cluster M5 collected from the literature. As a very important by-product of this comparison, we found that pulsation periods of BL stars follow Gaussian distribution with the mean period of $0.54\pm 0.07$\,d, while the modulation periods show log-normal distribution with centre at $\log(P_{\rm m}~{\rm [d]})=1.78\pm0.30$\,dex. This means that 99.7\,\% of all known modulated stars have BL~periods between 7.6 and 478 days. We discuss the identification of long modulation periods and show, that a significant percentage of stars showing long-term period variations could be classified as BL stars.}
   {}


\titlerunning{ASAS Blazhko stars and the distribution of modulation periods}
\authorrunning{Skarka et al.}

   \maketitle
%

\section{Introduction}
	
	Together with ultra-precise space measurements from the {\it Kepler} and {\it CoRoT} telescopes, the large-scale surveys play crucial role in investigation of long-term amplitude and frequency (phase) modulation in RR Lyrae stars known as the Blazhko (BL) effect \citep{blazhko1907}. Accurate space data helped uncover various dynamical processes present only in modulated stars (for example, period doubling \citep{szabo2010,kollath2011} and additional radial and non-radial modes \citep{benko2014,szabo2014b}). Data from the ground-based surveys, on the other hand, are very useful for studying large samples statistically and investigating general and common characteristics of modulated stars \citep[e.g.][]{alcock2003,soszynski2011}.
	
	Both approaches are important in their own way because the nature of the BL effect has not been fully explained yet. The true percentage of modulated stars is still unknown \citep[see e.g.][]{jurcsik2009,kovacs2015}. Even more interestingly, it is not clear why some stars show modulation and some with very similar physical properties do not \citep[e.g.][]{skarka2014b}. Knowledge about characteristics of the BL effect is still insufficient to give a reliable answer to the question of if and how the pulsation period is connected with the length and amplitude of the modulation. \citet{jurcsik2005a} showed that the modulation amplitude of short-pulsation-period RR Lyrae stars can be higher than for long-period stars. \citet{benko2014} and \citet{benko2015} suggest a monotonic dependence between the strength of amplitude modulation and the period of the BL~cycle (the longer the cycle, the larger the amplitude). While the upper limit of the modulation length is still unclear, the bottom limit is probably defined by the upper limit of the rotational velocity of a star. RR Lyrae stars with short pulsation periods can have short modulation periods, while long-pulsation-period stars can only have a BL cycle that is longer than about 20\,d \citep{jurcsik2005b}. 
	
Several studies report on the mode switching and (or) disappearing (reappearing) of the BL effect, suggesting that the phenomenon could be of temporal nature \citep[see e.g.][]{szeidl1976,sodor2007,goranskij2010,jurcsik2012,plachy2014}. Many stars also show variations of their modulation cycles \citep[see e.g.][]{guggenberger2012,benko2014,szabo2014b}. In some BL stars, for example in RR Lyrae itself, the modulation reappears cyclically during an additional four-year cycle \citep{detre1973,leborgne2014,poretti2016}. There are many more interesting features of the BL effect. Extensive reviews of the BL effect are provided in recent overviews by \citet{szabo2014a}, \citet{kovacs2015}, and \citet{smolec2016}.
		
	In his study based on the brightest RRab stars observed by the {\it All Sky Automated Survey} (ASAS) \citep[e.g.][]{pojmanski1997,pojmanski2001} and {\it the Super Wide Angle Search for Planets} \citep[SuperWASP;][]{pollacco2006}, \citet{skarka2014a}revealed the shortcomings of current automated procedures dedicated for the search for the BL effect and the need for permanent supervision of all steps when analysing noisy ground-based data dominated by various aliases. His study also showed that there are many undiscovered modulated stars even in previously studied data sets. Because frequency analysis and searching for the manifestations of the BL effect in many targets individually is a very time-consuming task, we establish a new group involving professionals, students, and amateur astronomers that aim to analyse data from various surveys. The project was called SERMON which is the abbreviation of {\it SEarch for Rr lyraes with MOdulatioN}. 
		
	In this paper we focus on RRab stars observed by ASAS that are brighter than 13.5\,mag in maximum light to investigate their frequency spectra trying to reveal the BL effect. Because the ASAS data have special characteristics and need special handling, we describe these data in detail in Sect. \ref{Sec:ASAS}. We also describe the sample selection in this section. Analysis, criteria for the detection of the BL effect and discussion about the sample stars are described in Sect. \ref{Sec:Analysis}. In Sect. \ref{Sec:PeriodDistribution} we investigate the distribution of modulation periods. The results are discussed in Sect. \ref{Sec:Discussion}, the content of the paper is summarized in Sect. \ref{Sec:Conclusions}.

\section{The ASAS data}\label{Sec:ASAS}

The ASAS-3 survey is performed by small-aperture telescopes that are primarily dedicated to search for new variables among stars with declination below $+28^{\circ}$ \citep[technical details can be found in][]{pojmanski2001}. The telescopes regularly scan the sky, and typically observe a particular object once per a few days with a three-minute exposure in $V$ filter.

The publicly available data typically contain several hundreds of data points, which are marked with flags from A (the best quality) to D (the worst quality). The photometry often consists of several blocks in the data files because each star is usually observed in a few different fields. Magnitudes with errors are provided in five apertures with diameter between 2 and 6 pixels. The time span of the data is about 3300 days in most cases.

\begin{figure}[htbp!]
	\begin{center}
		\includegraphics[width=\hsize]{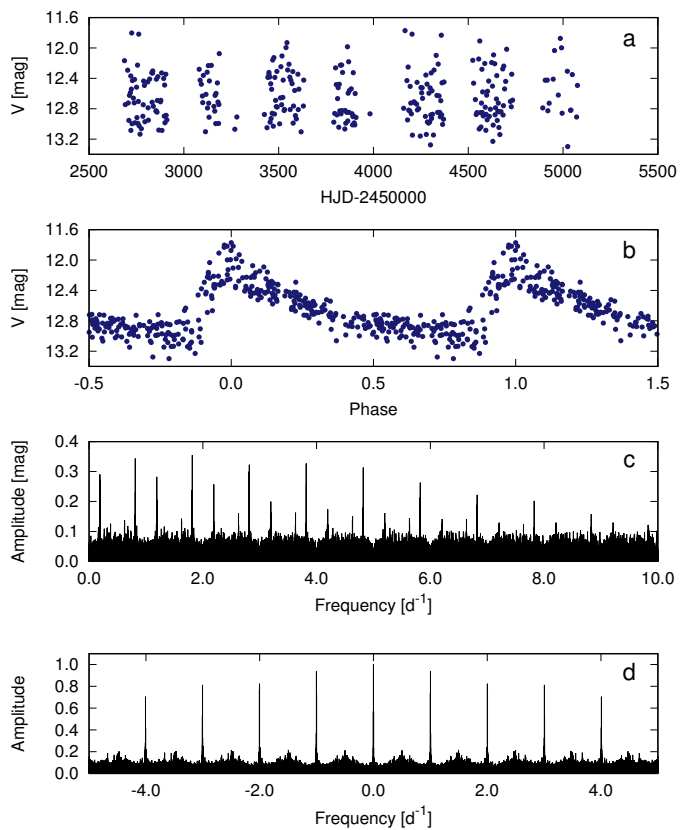}
		\caption{Typical ASAS data demonstrated on V1124 Her. Panels `a' and `b' show the data distribution and data phased with the basic pulsation period. Panel `c' shows corresponding frequency spectrum; panel `d' shows the spectral window.}\label{Fig:Data}	
	\end{center}
\end{figure}

The philosophy of the survey has a deep impact on data characteristics. Low sampling rate causes the frequency spectra to be strongly dominated by daily aliases (Fig. \ref{Fig:Data}). Because the angular resolution of the ASAS instruments is very low (about 15 arcsec/pixel), even with the smallest aperture the light of the target star could be influenced by neighbouring star(s), which is (are) closer than 30 arcsec. This fact could seriously bias the data. We discuss this problem in detail in Appendix \ref{Sec:App_Blends}. An extensive overview of the ASAS data characteristics can be found in \citet{pigulski2014}.  

\subsection{The sample selection and data cleaning}\label{Subsec:Sample}

This paper represents the extension of the study by \citet{skarka2014a} with fainter stars and stars with more scattered data. To some extend it complements the analysis by \citet{szczygiel2007} who performed an automatic search on the full sample of all types of RR Lyrae stars in the ASAS Catalogue of Variable Stars (ACVS). 

The selection process started with the the International Variable Star indeX catalogue \citep[VSX;][version from 27th October, 2014]{watson2006}, in which 1760 stars of RRab type with maximum light brighter than 13.5\,mag and declination below +28$^{\circ}$ were listed. The declination limit is given by the observational constraints of the ASAS survey. The magnitude limit was chosen according to several tests as the most limiting for reliable analysis because the scatter steeply increases with decreasing brightness. The corresponding photometric error of a data point at 13.5\,mag is about 0.15\,mag \citep{pojmanski2002}.

Stars analysed by \citet{skarka2014a} were ignored. Then we checked for the availability of the data via the online VSX tools. This resulted in 1435 stars that were observed by ASAS. After downloading the data, we obtained a basic idea about the quality of the data, checked for possible blends in the area with diameter of 30 arcsec, and made a note (col. Blend in Tables \ref{Tab:BlazhkoStars}, \ref{Tab:Known-detections}, \ref{Tab:OtherTypes}, and col. Rem in Table \ref{Tab:PC}). We used only data with flags A and B in aperture with the smallest average photometry error. Possible shifts in the mean magnitude between blocks were ignored because of the poor data sampling and problems with blends. All points that differed more than 1.5\,mag from the median value were automatically removed. No other automatic procedure was applied for improving the data to avoid possible misinterpretation because, for example, scatter caused by modulation could be very easily confused with observational scatter.

In the next step, all stars with less than 150 points were discarded. Then we performed a visual inspection of all light curves again, and searched for the basic pulsation period in stars where the value given in VSX was not reliable. In stars where suitable period was found (or was available in VSX) obvious outliers were manually deleted. Stars with extremely scattered data were removed from the sample. This kind of pre-analysis gave us an idea about some peculiarities in the objects and allowed us to identify duplicates (Sect. \ref{Subsec:Duplicates}) and stars of different variable types (Sect. \ref{Subsec:OtherVariables}). Together with these peculiar objects the sample for further analysis contained 1234 stars.

\section{Analysis and identification of the BL stars}\label{Sec:Analysis}

A mathematical description of simultaneous amplitude and phase (frequency) modulation suggests the rise of an additional peak with modulation frequency (\fm) and equidistant peaks near the basic pulsation frequency components \citep[\kf0$\pm$\fm; for details see][]{benko2011,szeidl2012}. This equidistant spacing is equivalent to \fm. Because the presence of the side peaks in the frequency spectra is the most distinct and convincing evidence of the BL~effect, which is also detectable in scattered data, we searched for this feature using \textsc{Period04} software \citep{lenz2005}. Owing to the characteristics of the data, it was necessary to pay special attention to properly identify correct peaks and avoid possible confusion with aliases. 

It is likely that variable contamination with parasite light from nearby stars entering the aperture could resemble the amplitude modulation in some of the sample stars. In such cases, the mean magnitude and total amplitude of the light variations varies in time and produces conspicuous peak in the low-frequency range of the frequency spectrum. A visual check of the data itself would not help much because of large scatter and poor sampling, and statistics fails too for the same reasons. Therefore, the only way to reveal this issue is to identify the false modulation peak. No such peak was observed in any of the sample stars, however, we cannot exclude that some of the stars could be influenced by this problem. This is one of the reasons why we give the information about blends in the tables.   

\begin{figure}[htbp!]
	\begin{center}
		\includegraphics[width=\hsize]{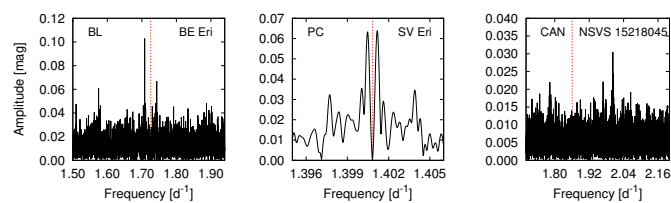}
		\caption{Typical frequency spectra after prewhitening with the basic pulsation components for BL stars, PC, and candidate stars (from left to the right). The position of the basic pulsation frequency is shown with the red dashed line.}\label{Fig:FreqSpec}	
	\end{center}
\end{figure}

A star was marked as a BL star when a peak with $S/N>3.5$ in the interval of $f_{0}\pm 0.2$\,c/d was identified. It was found by \citet{nagy2006} that in stars with frequency separation of less than 1.5/{\it TS} (periods longer than 2/3{\it TS}), where {\it TS} is the time span of the data, it is impossible to distinguish unambiguously between BL modulation and long-term continuous period change (instability) of the main pulsation period (i.e. period-change stars, PC class, as described in the next paragraph). Therefore, only peaks with separation larger than 1.5/{\it TS} were considered a consequence of the BL effect. An example of the frequency spectrum for a BL star is shown in the left-hand panel of Fig. \ref{Fig:FreqSpec}, where an equidistant triplet is seen.

The second category that we adopted is a PC group. Stars assigned to this group are usually those showing either unresolved peaks (peaks with frequency distance lower than 1/{\it TS}), or it is impossible to prewhiten the peaks in a few consecutive steps \citep{alcock2000,nagy2006}. Following our period limit for BL stars, each star with frequency separation lower than 1.5/{\it TS} was assigned to PC class as well. However, it is sometimes difficult to properly decide about BL/PC category, thus, we adopted these conservative criteria to make sure that PC stars do not contaminate significantly our BL sample (see also the discussion in Sect. \ref{Sec:Discussion}). A typical frequency spectrum of a PC star is shown in the middle panel of Fig. \ref{Fig:FreqSpec}.  

Finally, we use candidate category for stars that would belong to the BL group, but the peaks have amplitudes under the detection limit or very close to integer daily frequencies (1\,c/d, 2\,c/d etc., the right-hand panel of Fig. \ref{Fig:FreqSpec}).

\subsection{Stars with the BL effect}\label{Subsec:Blazhko}

An online version of the BlaSGalF database\footnote{\url{http://physics.muni.cz/~blasgalf/}} containing Galactic field RR Lyrae stars with the BL effect reported in literature contains 58 stars from our sample \citep[version from 29th February, 2016,][]{skarka2013}. We detected the side peak in only 39 of these stars (see Table \ref{Tab:Known-detections}). One of these 58 stars belongs to our candidate category, one was found to be of different variability type, and one star is a  PC star. In remaining 16 stars we did not detect any peak possibly related to modulation (see Table \ref{Tab:Nondetections}). 

Although we cannot be sure in 100 percent of the stars, we assume, that the nondetection of the BL effect is a consequence of the data characteristics rather then the result of absence of the BL effect. We found two reasons for that. First, the noise level of the residuals after removing significant frequencies is typically between 0.01 and 0.02\,mag, which means that sometimes side peaks with amplitude below $\sim 0.07$\,mag do not fulfil $S/N>3.5$. Second, some of the stars surely showed large-amplitude BL effect at the times when ASAS gathered the data. For example, \citet{jurcsik2009} easily detected the modulation in the frequency spectra of AQ Lyr and UZ Vir using high-quality photometric observations gathered with a 60-cm telescope between 2004 and 2009. They also did not detect any clear sign of modulation in ASAS and NSVS data that is confirmed by our analysis. In any case, the nondetection of the BL effect in ASAS data should not be considered as a proof of the absence of the modulation.


In 48 stars the BL effect was detected for the first time (see Table \ref{Tab:BlazhkoStars}). In addition to the information about pulsation and modulation periods resulting from our analysis (cols. 2 and 3) and corresponding errors in the last digits in parentheses, we give information about nearby stars closer than 30 arcsec (18 objects with `+' in col. 5 named Blend), and about the number of data points and the time span of the data  (cols. 6 \emph{N} and 7 \emph{TS}). Four stars denoted with `+' in the last column Rem, in which we detected additional, either unresolved, or unique peaks close to \kf0, are suspected of multiple modulation (category $\nu$M from \citet{alcock2003} and MC from \citet{nagy2006}), or additional long-period variation. Instead of amplitudes of the side peaks, which could be influenced by nearby stars, we give \emph{S/N} in the fourth column. Errors of periods (in all tables) are $2\sigma$ formal uncertainties from the least-squares fitting method.

\newcolumntype{L}{>{\footnotesize}l}
\newcolumntype{M}{>{\footnotesize}c}
\newcolumntype{R}{>{\footnotesize}r}
\setlength\tabcolsep{6pt}
\renewcommand{\arraystretch}{0.85}
\begin{table*}
		\centering
		\caption{Stars identified as modulated for the first time. Columns $P_{\rm Puls}$ and $P_{\rm m}$ give pulsation and modulation periods and corresponding $2\sigma$ errors in the last digits; column \emph{S/N} gives the signal-to-noise ratio; column Blend gives information about nearby stars; columns \emph{N} and \emph{TS} give the number of data points and the time span of the data. Stars denoted with `+' in column Rem are suspected of multiple modulation or additional long-period variation.}		
	\begin{tabular}{LMRMMMMM}\hline \hline 
ID&$P_{\rm Puls}$ [d]&$P_{\rm m}$ [d]&\emph{S/N}&Blend&\emph{N}&{\it TS} [d]&Rem\\ \hline
ASAS J103622-3722.1&0.468036(2)&86.2(4)&4.8&&553&3289&\\
CH Aps&0.509710(7)&730(40)&3.9&&616&3176&\\
ASAS J211737+0011.8&0.489135(4)&43.59(14)&4.0&+&295&3235&\\
OW Aqr&0.655187(2)&172(2)&4.3&&404&3254&\\
V0356 Aqr&0.554594(2)&41.65(6)&4.7&&440&3219&\\
V0522 Cen&0.623662(2)&2100(240)&3.9&&582&3187&\\
V0570 Cen&0.604507(8)&65.35(32)&3.6&&470&3170&\\
BW CMa&0.526804(2)&45.89(12)&6.7&+&530&3299&\\
ASAS J112830-2429.7&0.678271(2)&11.37(2)&3.9&+&840&3172&\\
NSV 1888&0.458739(2)&980(60)&4.2&&708&3248&\\
RT Equ&0.444761(10)&106.8(8)&4.2&&366&2921&+\\
BE Eri&0.579536(4)&57.38(12)&6.5&&380&3214&\\
WZ Gru&0.536766(4)&27.35(4)&4.7&&360&3261&\\
ASAS J175143+1708.4&0.435601(4)&6.67(2)&3.8&+&610&2404&\\
V1332 Her&0.522353(4)&21.59(4)&3.9&&435&2404&\\
ASAS J023020-5908.1&0.572857(4)&29.85(4)&5.3&&714&3279&\\
UW Hor&0.662156(14)&300(8)&5.2&&560&3297&\\
ASAS J094115-1348.8&0.467569(8)&101.0(6)&4.7&+&295&3141&\\
ASAS J104924-2812.5&0.574553(4)&93.3(6)&5.1&&571&3299&\\
ASAS J110733-2944.1&0.526282(4)&63.48(20)&5.1&&592&3154&\\
ASAS J145702-2642.6&0.586273(8)&75.26(36)&4.2&+&419&3170&\\
NSV 14546&0.349662(2)&1070(60)&4.4&&733&3294&\\
ASAS J151317-1445.9&0.613006(4)&43.33(8)&6.8&+&588&2965&\\
CU Lib&0.506956(6)&121.6(8)&4.6&+&473&3175&+\\
AC Men&0.553409(4)&120.2(8)&4.5&+&380&3288&\\
ASAS J055109-7219.4&0.558809(4)&168.3(16)&4.4&&634&3296&\\
ASAS J095835-8311.7&0.543469(2)&243(2)&8.4&&1344&3294&\\
BN Oct&0.508267(2)&23.93(4)&5.2&&580&3275&\\
EE Oct&0.444320(2)&175.9(16)&4.5&&750&3293&\\
XZ Oct&0.473867(2)&26.00(4)&4.9&&739&3251&\\
CSS J165728.3+055825&0.592147(2)&406(8)&5.9&&343&2959&\\
FU Pav&0.454554(6)&206(4)&4.0&&280&3215&\\
GV Pav&0.575562(4)&109.4(6)&5.2&&481&3215&\\
NSVS 11726192&0.576882(8)&31.61(6)&4.1&&388&3239&\\
SS Pic&0.494099(2)&61.86(14)&5.9&&621&3287&\\
RR Pyx&0.491078(4)&30.17(4)&5.2&+&878&3295&\\
AK Scl&0.494987(2)&94.2(6)&5.5&&558&3283&\\
CN Scl&0.585871(4)&1270(80)&4.2&&600&3286&+\\
V0484 Ser&0.511390(4)&119(1)&3.6&&324&2946&\\
ASAS J181510-3545.7&0.579472(4)&127.7(6)&5.2&+&1200&3174&\\
ASAS J193321-2517.6&0.507410(2)&55.7(2)&4.4&+&438&3132&+\\
V1069 Sgr&0.478929(6)&206(2)&5.3&+&501&3067&\\
V2281 Sgr&0.479455(6)&180(2)&3.8&+&360&3271&\\
ASAS J185332-5133.3&0.611209(8)&432(10)&3.9&&466&3155&\\
ASAS J162246-6723.5&0.543464(2)&85.7(4)&4.7&+&909&3181&\\
UX Tuc&0.509089(8)&50.25(16)&4.2&+&288&3254&\\
V0420 Vel&0.605077(2)&66.38(10)&10.2&+&598&3299&\\
ASAS J073358-6508.1&0.538119(6)&113.2(8)&3.9&+&510&3299&\\	  
 \hline
	\end{tabular}\label{Tab:BlazhkoStars}
	\end{table*}

\begin{figure*}
	\begin{center}
		\includegraphics[scale=0.9]{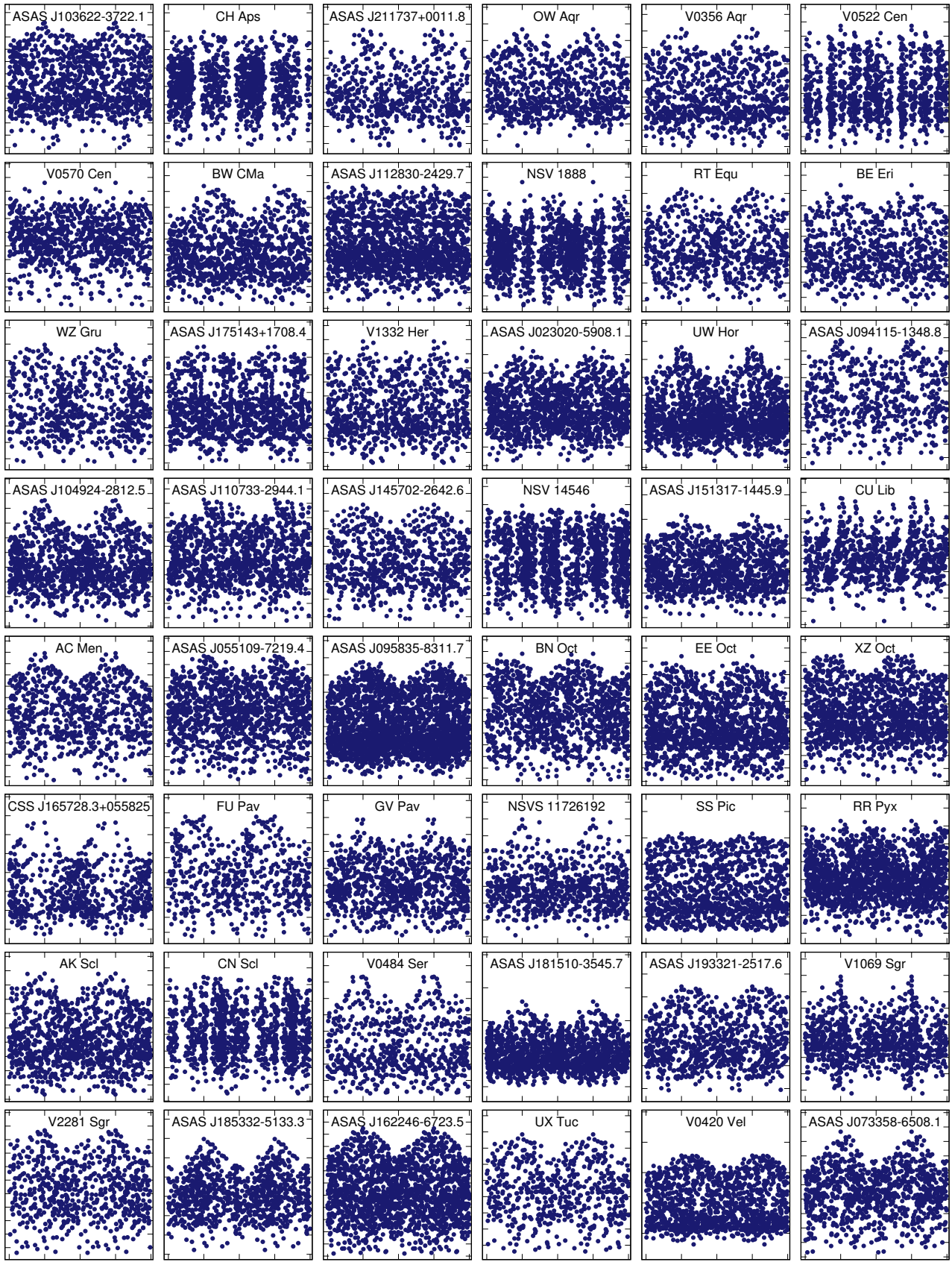}
		\caption{BL stars phased with modulation periods from Table \ref{Tab:BlazhkoStars}. As the zero epoch we simply took HJD of the brightest data point. Ticks at the ordinate show interval of 0.2\,mag, at abscissa the interval is 0.5 in BL phase.}\label{Fig:Mosaic}	
	\end{center}
\end{figure*}

All data of newly identified modulated stars phased with the modulation periods from Table \ref{Tab:BlazhkoStars} are shown in Fig. \ref{Fig:Mosaic}. In most of the stars the modulation envelope is well apparent. This figure also shows a variety of envelope shapes and modulation amplitudes. Some of the stars, for example ASAS J181510-3545.7 and V0420 Vel, have very small total amplitudes compared to the other stars, which is most likely a result of blends with nearby stars. 

\subsection{Stars with unresolved peaks}\label{Subsec:PC}

Our 22 PC stars are listed in Table \ref{Tab:PC}. The modulation periods suggested by the frequency spacing range from 1900\,d in BK Eri to 4000\,d in SV Eri. The latter is a star with the most rapid period change known among RR Lyrae stars so far \citep{poretti2016}. When the data are phased with the suggested modulation periods, DI Aps, QZ Cen, V0584 Cen, NSVS 14009, V0898 Mon, and V0687~Pup show apparent amplitude changes. It may be possible that in a data set with long enough time span, these stars will turn out to be classical BL stars with very long modulation periods (see also Sect. \ref{Sec:Discussion} and Fig. \ref{Fig:PC_stars}.)

Short pulsation periods of ASASJ 054843-1627.0, J160612-4319.2 and their low amplitude and phase Fourier coefficients (transformed to $I$-band filter) suggest that they are of RRc type when compared to characteristics of Galactic bulge RR Lyrae parameters provided by \citet{soszynski2011}.

\subsection{Candidates for BL stars}\label{Subsec:Candidates}

In seven stars we noticed peaks within $f_{0}\pm0.2$\,c/d, but we mark them as candidates from the reasons discussed previously. In YZ Aps, NSVS 15218045 (see Fig. \ref{Fig:FreqSpec}), ASAS J054238-2557.3, and V1211 Sgr, the suspicious peaks were very close to integer multiples of 1\,c/d, which is always problematic. In ASAS J081549-0732.9 and ASAS J194745-4539.7 only peaks with $S/N<3.5$ were detected.

\newcolumntype{C}{>{\footnotesize}c}
\setlength\tabcolsep{1pt}
\begin{table}[htbp]
		\centering
		\caption{Stars showing close, unresolved peaks near $f_{0}$ that were marked as PC. The notation of the columns is the same as in Table \ref{Tab:BlazhkoStars}. The star with `*' is known to the literature as a BL star.}		
	\begin{tabular}{LLCCCCC}\hline \hline 
ID&$P_{\rm Puls}$ [d]&$P_{\rm m}$ [d]&S/N&N&{\it TS} [d]&Rem\\ \hline
DI Aps&0.519187(6)&2700(300)&7.7&509&3212&\\
ASAS J204705-0919.2&0.508809(6)&2500(200)&6.7&369&3146&+\\
ASAS J083815-6025.9&0.465751(5)&2400(400)&5.9&550&3293&b\\
QZ Cen&0.502052(6)&2200(200)&6.1&634&3188&b\\
V0584 Cen&0.459032(2)&3200(600)&5.0&418&3153&\\
RS Crv&0.536834(6)&2200(200)&6.0&727&3169&\\
BK Eri&0.548148(4)&1900(100)&4.9&363&3243&+\\
SV Eri&0.713861(4)&2800(100)&10.0&536&3296&\\
NSV 14009&0.563985(6)&3000(200)&6.4&585&3256&\\
ASAS J054843-1627.0*&0.376709(4)&3000(200)&7.9&544&3299&b\\
PS Lup&0.471878(2)&3200(400)&5.7&663&3191&+\\
V0898 Mon&0.541493(12)&2200(200)&5.0&316&3078&b\\
ASAS J160612-4319.2&0.356944(2)&2800(200)&8.0&639&3194&b\\
ASAS J165834-8458.6&0.549312(6)&3200(500)&5.1&561&3241&\\
V0455 Oph&0.453893(2)&2200(400)&3.6&377&2586&b\\
KV Pav&0.527929(6)&4000(600)&6.9&528&3289&\\
V0338 Pav&0.530827(6)&2600(400)&4.4&330&3273&b\\
ASAS J075127-4136.3&0.335716(2)&3000(300)&6.8&610&3296&b\\
V0687 Pup&0.575194(8)&2500(300)&5.8&483&3246&b\\
ASAS J200147-2153.8&0.546631(14)&2500(400)&4.3&562&3144&b\\
ASAS J190713-5205.7&0.526724(2)&3100(600)&4.0&725&3173&\\
GR Tel&0.611916(4)&2900(400)&4.9&696&3134&b\\
 \hline
	\end{tabular}\label{Tab:PC}
	\tablefoot{Blended stars are marked with `b' and stars with additional peaks are marked with `+' in the column Rem.}
	\end{table}

\subsection{Duplicates}\label{Subsec:Duplicates}

During our analysis we identified three stars as duplicates because a star with a different ID is located almost at the same coordinates and it has similar pulsation (variability) properties. Actually, this was a trigger to perform a detail analysis of this problem regarding the total VSX catalogue \citep{liska2015}. Our duplicates are given in Appendix B, Table \ref{Tab:Duplicates}.

\subsection{Stars with different variable type}\label{Subsec:OtherVariables}

In 24 stars (Table \ref{Tab:Nonvariables} in Appendix B) we detected no signs of variability, 19 other stars are probably not RRab stars, but we suspect that they are of different variability type (Table \ref{Tab:OtherTypes}, Fig. \ref{Fig:Others} in Appendix B). We give suggested variability types that were estimated on the basis of period, light-curve shape, and $B-V$ (Table \ref{Tab:OtherTypes}, Fig. \ref{Fig:Others}).

\subsection{Characteristics of sample BL stars}\label{Subsec:BlazhkoCharacteristics}

\begin{figure}[htbp!]
	\begin{center}
		\includegraphics[width=\hsize]{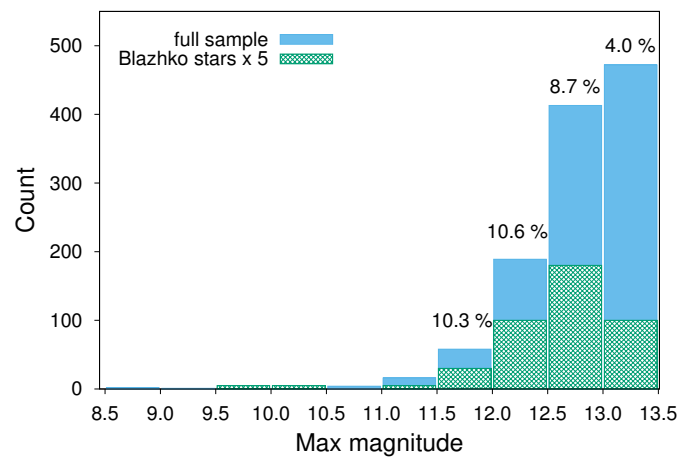}
		\caption{Magnitude distribution of all sample stars (blue) and BL stars (green grid). The distribution of modulated stars was enlarged five times for better visibility.}\label{Fig:Mag_dist}	
	\end{center}
\end{figure}

The percentage of all BL~stars in our sample (both known + new detections) is very low, at slightly less than 7\,\%. The reason is a large scatter that could hider the modulation, rather than a real lack of modulated stars among faint stars. Therefore, this number has nothing to do with the real amount of modulated stars, but only gives information about the detectability of modulation in the data set. Our assumption is reinforced by the look of the magnitude distribution of the sample stars shown in Fig. \ref{Fig:Mag_dist} (see the percentages above bins). The highest number of BL stars was detected between 12.5 and 13.0\,mag\,\footnote{The percentage is not the highest in this magnitude range.} and decreases steeply in the most populated area with faint stars with scattered data.

In the majority of the new BL stars the modulation is very apparent causing large amplitude changes (e.g. BW CMa, V0484 Ser), while in some of the stars the modulation in amplitude is barely seen (e.g. NSV 14546, SS Pic, see Fig. \ref{Fig:Mosaic}). Even in scattered ASAS data, a wide palette of the shapes of modulation envelopes is apparent. In this sense, one of the most interesting examples is UW Hor, which seems to show non-sinusoidal amplitude modulation. On the other hand, BN Oct and ASAS J162246-6723.5 show nearly sinusoidal variations in amplitude. The BL~peak \fm~itself was detected only in BL Col, a known BL~star. In four new BL stars and three PC stars we detected additional peaks, which could be interpreted as additional modulation, or signs of additional period change. Unfortunately, the low quality of the data does not allow us to investigate these peculiarities in detail.     

\section{Modulation-period distribution}\label{Sec:PeriodDistribution}

To get an idea about the new BL stars, we collected a sample of 1547 fundamental mode stars with estimated modulation period from the literature. This sample contains RRab stars from the Galactic field \citep[246 stars from the BlaSGalF database;][]{skarka2013}, the LMC \citep[731 stars;][]{alcock2003}, the Galactic bulge \citep[19+526 stars;][]{moskalik2003,collinge2006}, and globular cluster M5 \citep[25 stars;][]{jurcsik2011}. We considered all periods in stars with multiple modulation. For example, from RS Boo we have three values on our list. Together with our new detections (48), it amounts 1628 entries, which is, to our knowledge, the most extended sample of modulation periods investigated so far. 

\begin{figure}[htbp!]
	\begin{center}
		\includegraphics[width=\hsize]{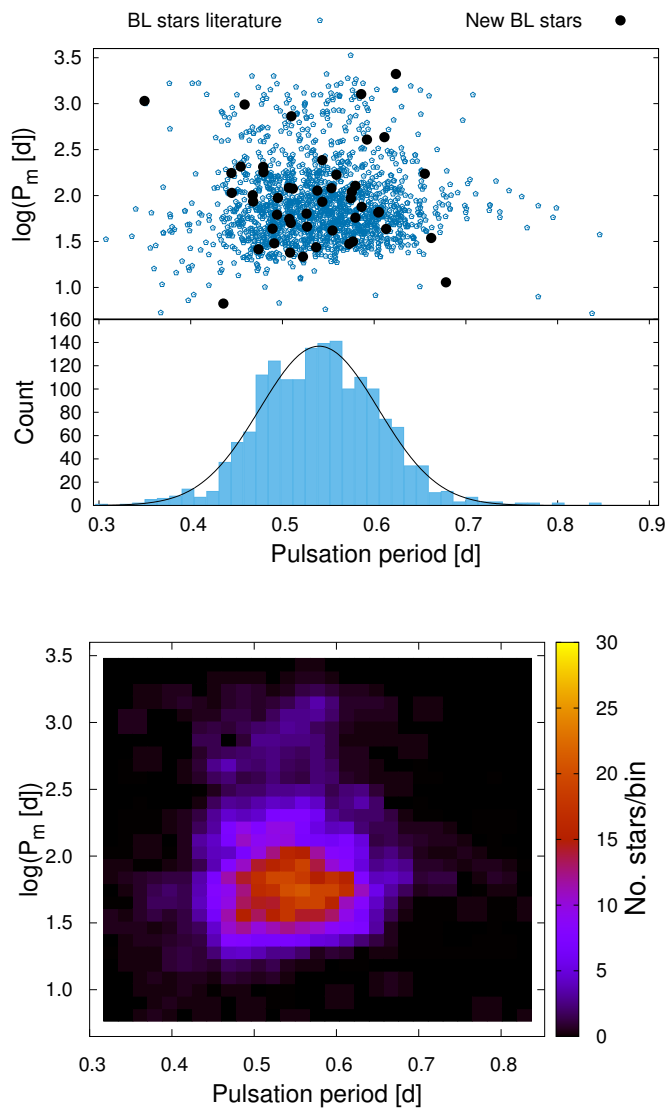}
		\caption{BL~period as a function of the pulsation period for known and new modulated stars (top), and corresponding density plot (bottom). The bottom part of the top panel shows the distribution of pulsation periods.}\label{Fig:Period_distribution}	
	\end{center}
\end{figure}

The modulation periods of newly identified BL stars (black circles in the top panel of Fig. \ref{Fig:Period_distribution}) match the distribution very well, and no sample star has extreme modulation period\footnote{The shortest BL~period was detected in ASAS J175143+1708.4 (6.67\,d), the longest in V0522 Cen (2100\,d), which is exactly the limit for BL stars that we adopted (2/3\emph{TS}).}. Two of our stars with the shortest modulation period, ASAS J175143+1708.4 and ASAS J112830-2429.7, however, seem to significantly differ from the general trend. The BL effect in these two stars should be proved independently. Light curve shape, pulsation period of only about 0.35\,d, and its Fourier parameters suggest that NSV 14546 could be an RRc-type rather than RRab-type star.   

To test the similarity of our sample with the sample of stars from literature, we performed the two-sample Kolmogorov-Smirnov test. The result is, that the agreement between the two populations cannot be ruled out with probability of 7\,\%. Thus we can conclude that our sample stars have a similar distribution of modulation periods as BL stars from literature (see the cumulative-distribution functions in the right-hand panel of Fig.~\ref{Fig:Histograms}).

However, the modulation-period distribution in Fig. \ref{Fig:Period_distribution} itself is far more interesting than the distribution of new BL stars\footnote{From here on we do not distinguish between new and known BL stars; all stars create one sample.}. While pulsation periods show a distribution close to the normal Gaussian distribution (the bottom part of the top panel in Fig. \ref{Fig:Period_distribution}), modulation periods follow the log-normal distribution. In other words, they show normal distribution when the BL~periods are plotted in logarithmic scale, which is nicely seen from the bottom panel of Fig. \ref{Fig:Period_distribution} and histograms in the left-hand panel of Fig.~\ref{Fig:Histograms}. The mean pulsation period of BL stars is $0.54\pm0.07$\,d, and the mean modulation period based on the log-normal distribution is $1.78\pm0.30$\,dex. This means that 99.7\,\% ($3\sigma$ interval) of all BL stars have modulation periods between 7.6 and 478 days.

\section{Discussion}\label{Sec:Discussion}

The log-normal distribution commonly appears in measured quantities that cannot be negative. Many of natural, social, and technical systems and quantities show such distribution, for example the length of a latent infection (from infection to burst of the disease), species abundance, age of marriage, and probability of failure of mechanical devices \citep[nice examples across the disciplines can be found in][]{limpert2001}. The question is why the pulsation periods have Gaussian distribution, while modulation periods show log-normal distribution. Why do the majority of BL stars have modulation periods between 30 and 120 days ($1\sigma$ interval)? At the present time, it is impossible to answer this question because no correlation between any of the physical characteristics and the length of the modulation period has been found.

\begin{figure*}
	\begin{center}
		\includegraphics[width=\hsize]{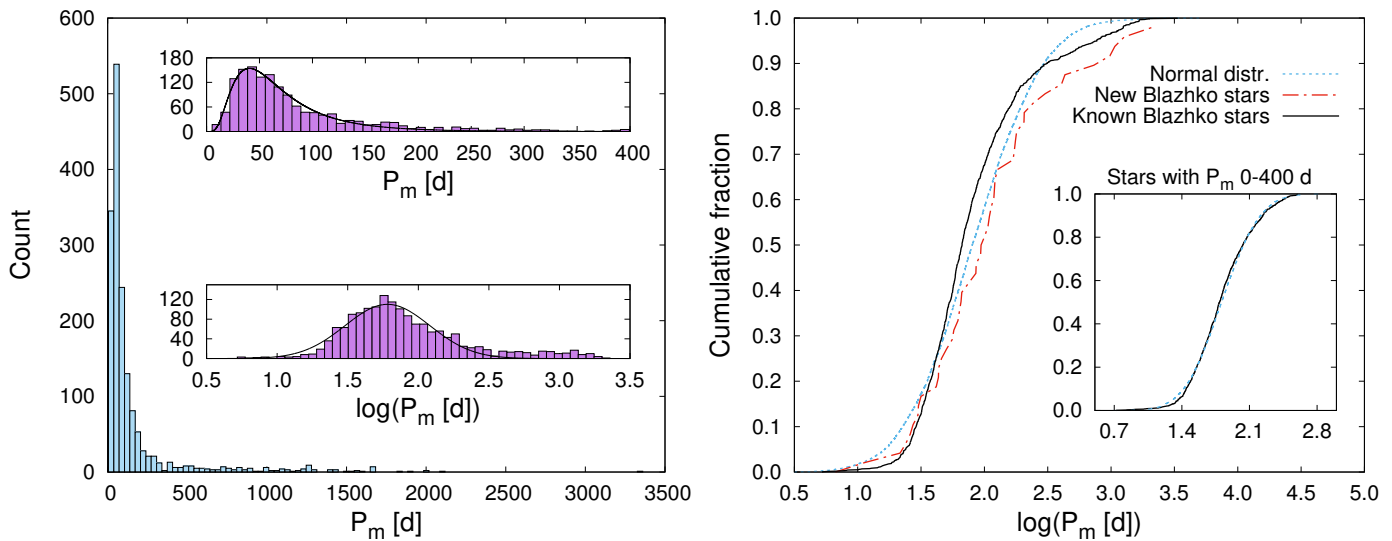}
		\caption{Distribution of modulation periods (left-hand panel) and cumulative distribution functions (right-hand panel). It is apparent that the overall distribution follows the log-normal distribution. The agreement with the normal distribution for stars with modulation periods between 0 and 400\,days shown in the insert in the right-hand panel is very good.}\label{Fig:Histograms}	
	\end{center}
\end{figure*}

The selection effects only have minor impact on the distribution, if any. The majority of used BL stars come from the data sets with time span that is several years long, where the modulation periods of several hundreds of days should be safely detectable. Yearly aliases cause problems with identification only in a very small part of the distribution (see the gap around 350\,d in the top insert in the left-hand panel of Fig. \ref{Fig:Histograms}). Thus we are convinced that the log-normal distribution is real. The very good agreement between the cumulative-distribution function for the periods in this interval and the normal distribution with the same parameters also confirms our assumption (the insert in the right-hand panel of Fig. \ref{Fig:Histograms}).

It seems that the log-normal distribution only represents the short-period part of the modulation period distribution well (0--400\,d), which has a long tail towards longer modulation periods. However, the distribution of the longest BL periods, which is probably much more numerous than our collection shows, is somewhat uncertain from several reasons. First, the entire sample is sharply cut at BL periods of about $\log P_{\rm m}\sim3.5$ ($\sim 3300$\,d). This limit is given by the extension of the longest data used for modulation period estimation (the ASAS-3 data with maximum time span of about 3300 days). Data from MACHO survey has time span of about 7.5 years. It can be naturally expected that the distribution continues to even longer modulation periods because we do not know the upper limit. What is probably the longest BL effect with period longer than 25 years is reported by \citet{jurcsik2016} in V144 in globular cluster~M3. 

\begin{figure}[htbp]
	\begin{center}
		\includegraphics[width=\hsize]{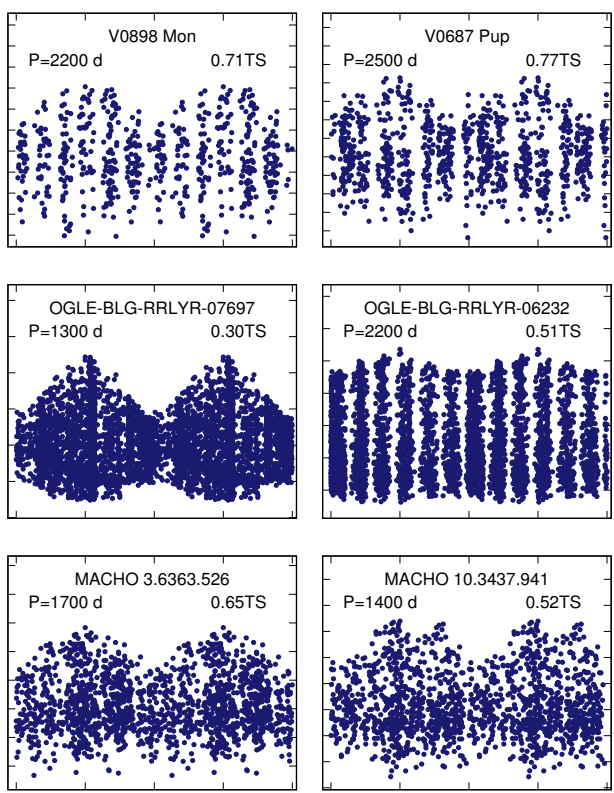}
		\caption{Stars originally classified as pure PC. There are stars from our sample in the top panel. The middle panels show OGLE-III data demonstrating that in long-enough data sets (original classification comes from OGLE-II data with three-times shorter time span) the stars could be classified as BL stars, not pure PC stars. The bottom panels show two stars that clearly show amplitude modulation with periods shorter than 2/3\emph{TS} suggesting classification as BL+PC, not pure~PC. The scales of the axes are the same as in Fig. \ref{Fig:Mosaic}.}\label{Fig:PC_stars}	
	\end{center}
\end{figure}

Recent observational investigations \citep[e.g.][]{benko2014} showed that the BL stars always show both amplitude and phase variations. The BL stars could then be identified through the variations in amplitude even when the modulation period is comparable with \emph{TS}, or longer. For example, V0898 Mon and V0687 Pup from our sample (top panels of Fig. \ref{Fig:PC_stars}) show significant amplitude changes with suggested periods that are a bit longer than 2/3\emph{TS}. Although the modulation periods are only very roughly estimated in these stars, the BL-phased light curves suggest, that they could safely be considered BL stars.

Another uncertainty relies in the problems with the BL/PC classification. Although secular period changes and nonstationarity in pulsation period are quite common among RR Lyrae stars \citep[see e.g.][]{leborgne2007}, we can expect a significant number of PC stars to be classified as BL stars in long-enough data sets. This is, for example, the case of OGLE-BLG-RRLYR-07697 and OGLE-BLG-RRLYR-06232. These two stars were classified as pure PC stars by \citet{collinge2006} on the basis of the four-year-long OGLE-II data. In OGLE-III data \citep{soszynski2011}, with a time span of about 12 years, these two stars appeared to be clear BL stars showing significant amplitude modulation with periods of 1300 and 2200\,d (the middle panels of Fig. \ref{Fig:PC_stars}). In OGLE-BLG-RRLYR-07697 the side peaks were easily prewhitenable\footnote{The star shows additional BL modulation with period of 36.48\,d.}. In OGLE-BLG-RRLYR-06232 we detected additional, unresolved structures very close to $f_{0}$, which cannot be fully eliminated, thus, suggesting some additional period variations. According to the criteria adopted in Sect. \ref{Sec:Analysis}, this star should be classified as both BL and PC star.

Further, we also suppose that stars showing dense patterns close to $f_{0}$, which cannot be prewhitened in a few steps, do not necessarily exclude the possibility of being BL stars. In the bottom panel of Fig. \ref{Fig:PC_stars} we see two stars identified as pure PC stars on the basis of MACHO data \citep{alcock2003}. Both of these stars clearly show large amplitude modulation with periods shorter than 2/3\emph{TS}. Thus, there is no reason why we do not assign them to the BL class with a possible additional PC component. It would be advisable to strictly distinguish between unresolved peaks close to $f_{0}$, which could mean instability of the main pulsation period, and other unresolved peaks near the side peaks, which in turn could mean the nonstationarity of the modulation itself, additional modulation, or some artificially generated periodicity in the data. In such cases the star should be classified as BL star with additional PC, rather then pure PC. 

As many previous investigations show \citep{lacluyze2004,sodor2007,detre1973,szeidl1976,leborgne2014,guggenberger2012}, the BL modulation itself can change on year-long time scale. Although the ASAS and MACHO data sets span rather long time bases, they typically do not contain sufficient number of data points to investigate such details. Therefore, many stars with changing BL modulation might remain undetected or classified as PC stars. These very few examples show that the distribution of modulation periods that are longer than $\sim1000$\,d is certainly more populated than we see in Fig. \ref{Fig:Histograms}.

\section{Summary and conclusions}\label{Sec:Conclusions}

We performed an in-depth study of frequency spectra of 1234 fundamental RR Lyrae stars observed by the ASAS survey, placing emphasis on proper identification of possible modulation. Therefore, we investigated each target individually within a new project SERMON employing professionals, amateurs, and students. We omitted stars with a low number of points ($<150$) and scattered light curves with insufficient quality for period analysis. The limiting magnitude was set to 13.5 in maximum light because of large scatter of the data in faint stars.

The criteria for accepting a star as modulated were purely on the basis of the appearance of the frequency spectra. In this sense we establish three groups: {\it i)} BL stars: stars with peak with \emph{S/N}$>3.5$ in the vicinity of $f_{0}$, no closer than 1/\emph{TS}, but closer than 0.2\,c/d; {\it ii)} BL~candidates: stars with one suspicious peak at integer daily frequencies, or with peaks close to $f_{0}$ with \emph{S/N}$<3.5$; and {\it iii)} PC: stars with unresolved side peaks or with a separation of the peaks suggesting modulation periods that are longer than 2/3\emph{TS}. Altogether we detected 87 BL stars, 48 of these stars for the first time. Even in low-cadence ASAS data with large scatter we observed a variety of possible shapes of modulation envelopes (Fig. \ref{Fig:Mosaic}). We also detected 7 candidates and 22 PC stars. When we put together stars from the ASAS-3 survey analysed by \citet{skarka2014a} with the sample analysed in this study, the detection efficiency of individual approach is about 12\,\% in ASAS data, which is more than two times better than in \citet{szczygiel2007}\footnote{It is worth noting that they analysed data with a time span two years shorter than our data sample.}.

For the comparison of newly identified BL stars, we collected a sample comprising 1547 stars with known BL periods from literature. The distribution of modulation periods of newly identified stars corresponds well to the distribution of periods of known BL stars. As a by-product of this analysis, we noticed that the pulsation periods of BL stars follow Gaussian distribution (mean $0.54\pm0.07$\,d), while the modulation periods are distributed log-normally with mean $\log P_{\rm m} {\rm [d]}=1.78\pm0.30$\,dex. Especially modulation periods in range 0--400\,d follow the log-normal distribution well. 

Behind its log-normal part ($P_{\rm m}>480$\,d), the distribution shows an almost constant number of stars per bin, however, this could be simply observational bias caused by the lack of suitable data with long-enough time span. Additional ambiguity comes from the difficulties in properly assigning a star to the PC or BL class. From our discussion about long-term modulation, it follows that stars showing amplitude modulation with period shorter than 2/3\emph{TS} can be safely classified as BL stars, no matter whether there are additional peaks that cannot be prewhitened. These unresolved peaks simply mean that some additional nonstationarity is present in the star, not that the star is unmodulated. Therefore, the classification criteria for PC stars could be revised and summarized in two simple points: \emph{i)} the separation of the side peaks and $f_{0}$ is lower than 1.5/\emph{TS}, whether or not the peaks are prewhitenable and \emph{ii)} the star shows no apparent amplitude changes. Combined data sets, for example OGLE-III and OGLE-IV, could shed more light on the problem with PC stars and allow us to establish more firm criteria.

The reasons standing behind the log-normal distribution are, unfortunately, unclear. Could there be any connection between the length of modulation period and evolutionary stadium, metallicity, and Oosterhoff groups? These questions would need a deep-in detailed investigation, which is out of scope of our study.

\section*{Acknowledgements}
We are very grateful to Johanna Jurcsik and G\'{e}za Kov\'{a}cs for their very useful comments on the first version of the manuscript. We also thank the referee for his/her useful suggestions. The financial support of the Hungarian National Research, Development and Innovation Office -- NKFIH K-115709 is acknowledged. M.\,S. acknowledges the support of the postdoctoral fellowship programme of the Hungarian Academy of Sciences at the Konkoly Observatory as a host institution. In addition to the mentioned grant, \'A.\,S. acknowledges financial support of the OTKA K-113117 grants, and the J\'anos Bolyai Research Scholarship of the Hungarian Academy of Sciences. This research has made use of the International Variable Star Index (VSX) database, operated at AAVSO, Cambridge, Massachusetts, USA, SIMBAD and VizieR catalogue databases, operated at CDS, Strasbourg, France, and NASA's Astrophysics Data System Bibliographic Services.

\appendix

\section{Blend troubles}\label{Sec:App_Blends}

\begin{figure}[htbp]
	\begin{center}
		\includegraphics[scale=0.5]{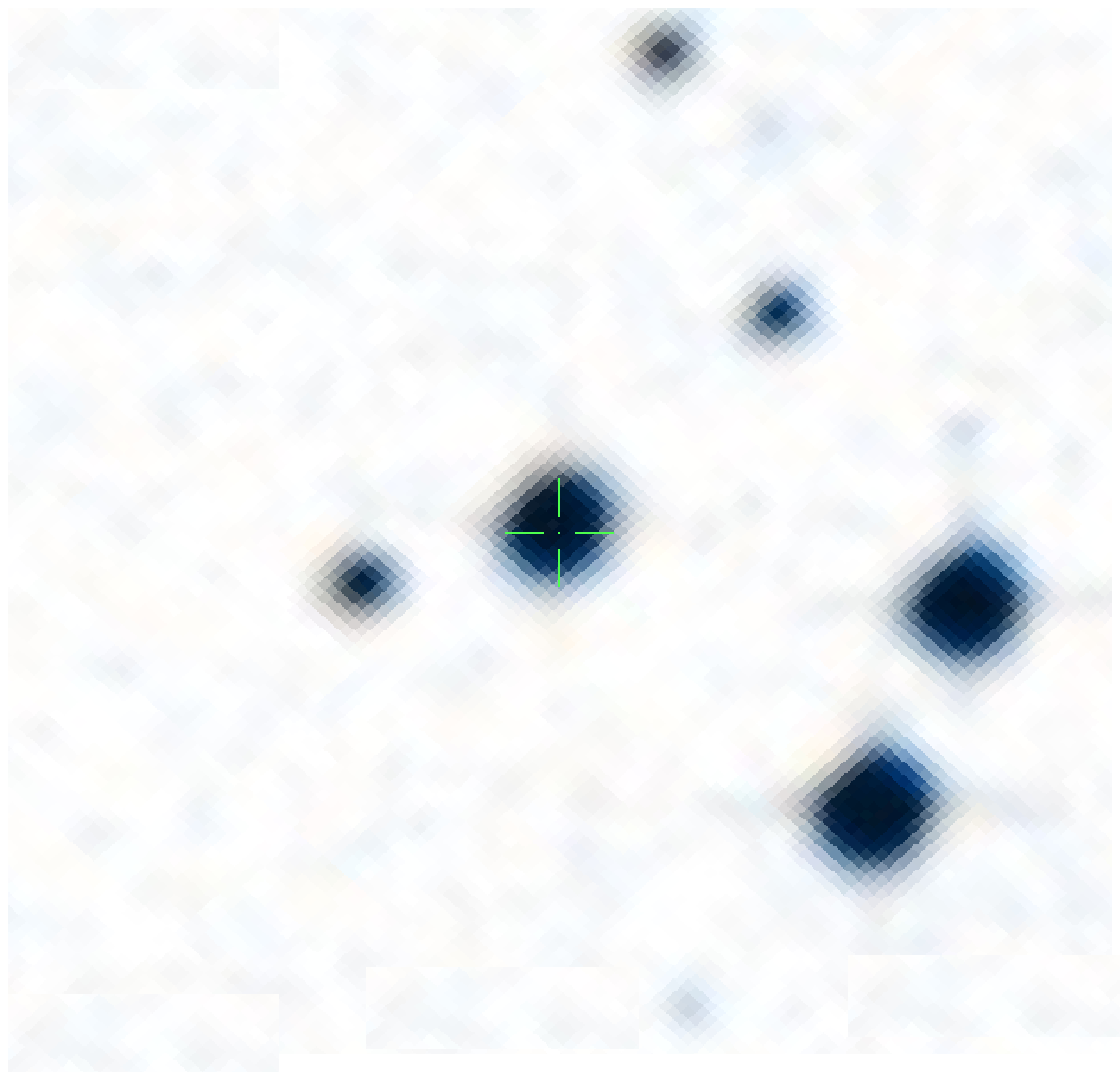}
		\includegraphics[width=\hsize]{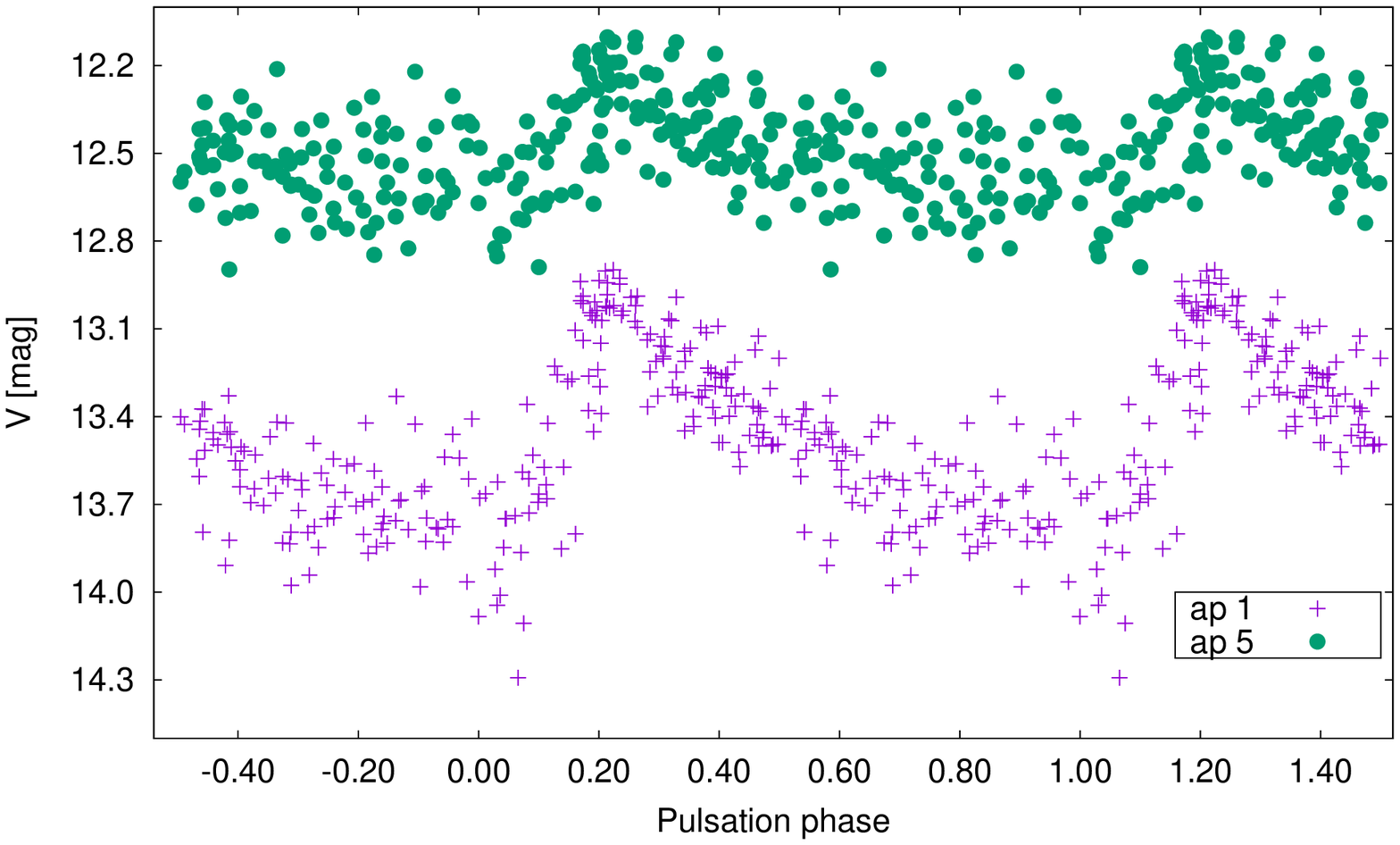}\\
		\includegraphics[width=\hsize]{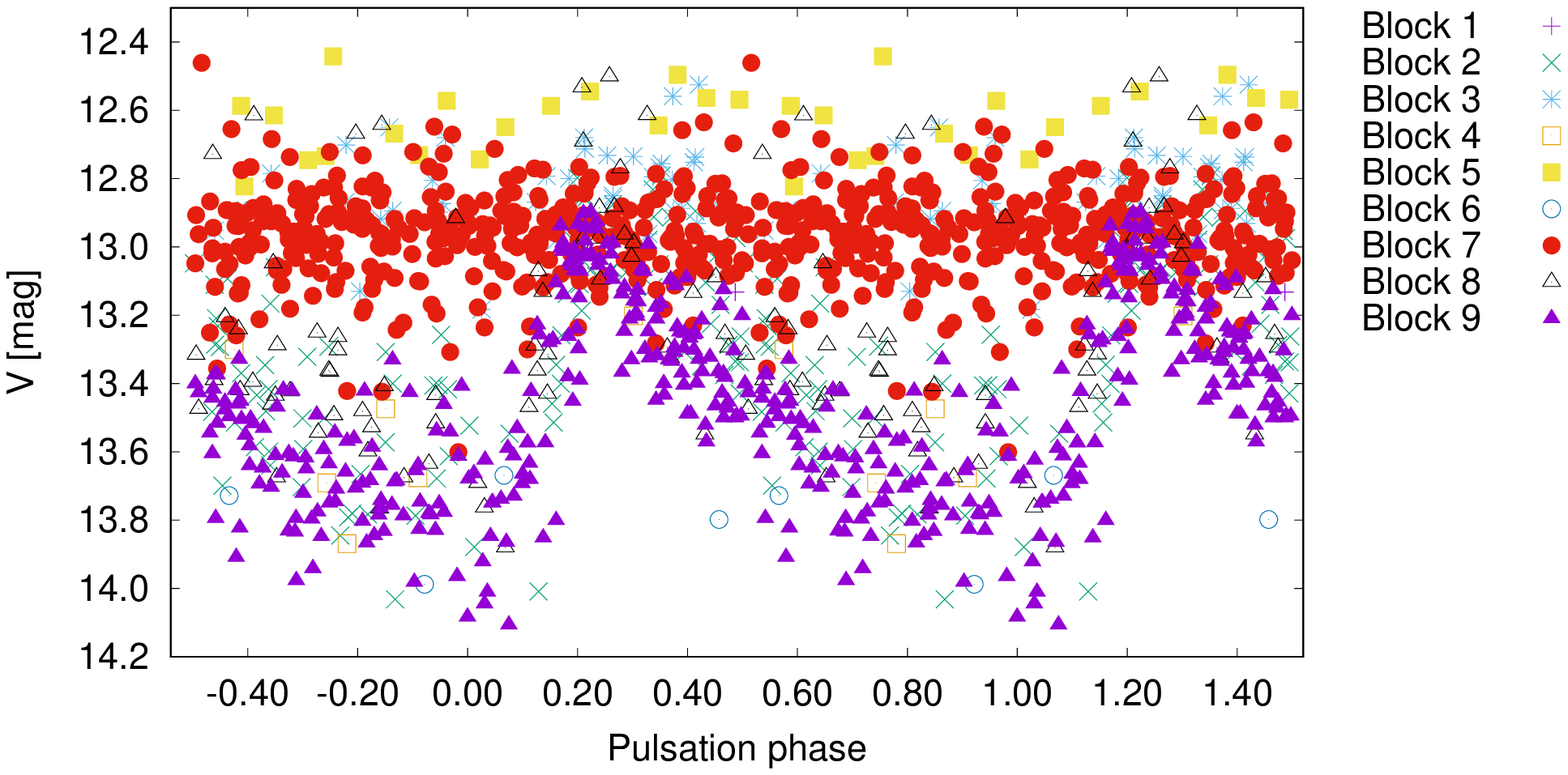}
		\caption{ASAS data of CSS J054243.2-114742 (star in the centre of the top left panel) phased with elements given in VSX. The top right panel demonstrates problems caused by large aperture, the bottom panel shows confusing the target with nearby stars in different fields (blocks of data).}\label{Fig:CSS}	
	\end{center}
\end{figure}

We demonstrate the problem with blends on CSS J054243.2-114742, where three similarly bright stars and three additional fainter stars are present within area with diameter of 1.5\,arcmin = 6 pixels in ASAS (the top panel of Fig. \ref{Fig:CSS}). The data set of this star comprises nine blocks of data points in ASAS-3 database. The middle panel of Fig. \ref{Fig:CSS} shows how the data from block 9 differs in apertures 1 and 5. It is obvious that in the largest aperture the light contribution of all neighbouring stars is taken into account. This results in higher brightness and lower pulsation amplitude in aperture 5 (6 pixels) than in aperture 1 (2 pixels).

An additional problem caused by close stars could arise from the combination of different fields. The bottom panel of Fig. \ref{Fig:CSS} shows the measurements in aperture 1 for different blocks of the data set. It is apparent that in different fields (blocks 3, 5, 7) the variable star was mistaken for one of the close nonvariable stars. Only blocks 2 and 9 probably show the data for the target star exclusively.

\section{Additional tables and figures}\label{Sec:App_Candidates}
	
\setlength\tabcolsep{4pt}
\renewcommand{\arraystretch}{0.9}
\begin{table}[htbp]
		\centering
		\caption{Known BL stars where we did not detect the modulation. The column \emph{S/N} gives information about \emph{S/N} of the highest peak detected after removing significant pulsation components in the vicinity of $f_{0}\pm0.2$\,c/d.}		
	\begin{tabular}{LLLLL}\hline \hline 
ID&$P_{\rm Puls}$ [d]&\emph{S/N}&$P_{\rm literature}$ [d]&Ref\\ \hline
AL Eri&0.656947(4)&3.3&-&K05\\
ASAS J030534-3116.1&0.496453(2)&3.2&6.67&SF07\\
V0461 Her&0.513006(6)&3.0&-&B14\\
V1162 Her&0.547929(6)&3.3&-&B14\\
GL Hya&0.505932(4)&3.4&157&B14\\
V0487 Hya&0.561482(6)&3.2&64.4&B14\\
ASAS J101200+1921.9&0.482830(10)&3.3&1141.03&SF07\\
MS Lib&0.441446(8)&3.1&105&B14\\
AQ Lyr&0.356809(4)&3.2&64.9&J09\\
V0829 Oph&0.569222(8)&3.1&165&Wa06\\
V2683 Oph&0.597012(4)&2.9&26.28&SF07\\
V0606 Peg&0.529658(4)&2.9&26.7&B14\\
FI Psc&0.531285(6)&3.2&>170&B14\\
CzeV0245&0.513328(4)&3.0&-&K12\\
UZ Vir&0.459394(4)&3.3&68.24&J09,S12\\
ASAS J194502+2434.2&0.845945(22)&3.2&37.56&SF07\\
 \hline
	\end{tabular}\label{Tab:Nondetections}
		\tablefoot{Ref: K05 -- \citet{kovacs2005}; SF07 -- \citet{szczygiel2007}; B14 -- \citet{bramich2014}; J09 – \citet{jurcsik2009}; Wa06 -- \citet{watson2006}; K12 -- \citet{kocian2012}; S12 -- \citet{sodor2012}.}
	\end{table}

\setlength\tabcolsep{4pt}
\renewcommand{\arraystretch}{0.9}
\begin{table}
		\centering
		\caption{Stars in which we cannot find any sign of variability or detect any period.}		
	\begin{tabular}{LL}\hline \hline 
ID	&	ID	\\ \hline
2MASS J06533015+1531072	&	NSVS 9525696	\\
CSS J163942.1+223202	&	V0429 Ori	\\
CSS J102417.2+161949	&	CSS J225023.1+171307	\\
CSS J150035.2-165444	&	ASAS J004535+0526.1	\\
CSS J150917.1-170252	&	MM Pup	\\
CSS J153512.2-145339	&	NSVS 19123091	\\
NSVS 12552884	&	NSVS 19168136	\\
NSVS 15305366	&	V0713 Sco	\\
CSS J164808.5+012515	&	ASAS J185602-1934.6	\\
OGLE BLG-RRLYR-00252	&	NSVS 19562915	\\
V0771 Oph	&	OGLE BLG-RRLYR-16496	\\
CSS J053340.7-021532	&	V0701 Sgr	\\
 \hline
	\end{tabular}\label{Tab:Nonvariables}
	\end{table}

\setlength\tabcolsep{12pt}
\renewcommand{\arraystretch}{0.9}
\begin{table}[htbp!]
		\centering
		\caption{Stars with duplicate occurrence in VSX (the same coordinates and pulsation properties).}		
	\begin{tabular}{LL}\hline \hline 
ID & Coord. (J2000) \\ \hline
ASAS J165728+0558.4 & 16 57 28.42 +05 58 25.5 \\\vspace{2mm}
\hspace{-0.8mm}CSS J165728.3+055825 & 16 57 28.39 +05 58 25.7 \\
CSS J160717.1-063352 & 16 07 17.12 $-$06 33 53.0 \\\vspace{2mm}
\hspace{-0.8mm}V0681 Oph & 16 07 15.33 $-$06 34 18.1 \\
CSS J213502.5+195641 & 21 35 02.59 +19 56 41.7 \\
\hspace{-0.2mm}ASAS J213502+1956.7 & 21 35 02.61 +19 56 41.0\\
 \hline
	\end{tabular}\label{Tab:Duplicates}
	\end{table}

\setlength\tabcolsep{4pt}
\renewcommand{\arraystretch}{0.9}
\begin{table*}
		\centering
		\caption{Known BL stars from literature detected in our ASAS sample. The notation of the columns is the same as in Table \ref{Tab:BlazhkoStars}.}		
	\begin{tabular}{LLLMMLLLL}\hline \hline 
ID&$P_{\rm Puls}$ [d]&$P_{\rm m}$ [d]&\emph{S/N}&Blend &\emph{N}&\emph{TS} [d]&$P_{\rm literature}$ [d]& Ref\\ \hline
CK Aps&0.623640(5)&56.4(2)&3.9&+&887&3205&-&K05\\
KM Aql&0.438191(4)&192(2)&4.7&&384&3162&192.2&SF07\\
PQ Aqr&0.512292(4)&120.3(9)&4.5&&362&3144&-&B14\\
V0354 Aqr&0.529555(6)&182(1)&6.2&&398&3291&181.2&SF07\\
ASAS J172721-5305.9&0.435434(2)&58.5(2)&4.8&&615&3171&58.66&SF07\\
ASAS J123812-4422.5&0.523543(6)&1540(120)&6.6&+&534&3162&1307.7&SF07\\
ASAS J135813-4215.1&0.523179(3)&145(1)&5.3&+&492&3197&146.01&SF07\\
GW Cet&0.516652(4)&84.2(4)&4.1&&462&3278&84.99&SF07\\
ASAS J053830-3554.4=BL Col&0.590761(4)&40.53(6)&6.3&&583&3289&40.05&K11\\
AI Crt&0.502901(4)&63.0(2)&5.8&&622&3176&63&WS05\\
V0365 Her&0.613178(10)&40.8(1)&4.5&+&224&2396&40&W06\\
V1124 Her&0.551018(4)&38.92(6)&6.4&+&296&2388&38.8&B14\\
RV Hor&0.572496(4)&79.7(3)&5.8&+&512&3264&79.81&SF07\\
ASAS J093731-1816.2&0.520920(8)&87.7(3)&6.0&&320&3294&87.73&SF07\\
ASAS J141025-2244.8&0.639874(4)&1700(100)&7.3&&567&3186&1556.66&SF07\\
V0430 Hya&0.496828(2)&58.5(3)&4.6&&363&3109&56.6&B14\\
V0486 Hya&0.508644(4)&53.4(2)&4.0&&405&3286&18.5&B14\\
V0543 Hya&0.598263(6)&58.8(2)&5.2&&668&3299&59&B14\\
CZ Ind&0.605147(4)&133(1)&4.1&&462&3257&133.38&SF07\\
V0550 Peg&0.493047(13)&21.98(4)&4.0&&219&3235&348.58&SF07\\
V1820 Ori&0.479040(6)&27.95(2)&6.5&&364&2625&27.917&dP13\\
ASAS J185719-6321.4&0.412008(2)&63.0(2)&4.6&&496&3149&61.39&SF07\\
CS Phe&0.484395(4)&61.2(2)&5.9&&530&3287&62.5&WS05\\
NSV 0539 = DR Dor&0.460421(2)&40.10(4)&7.0&&515&3256&40.17&SF07\\
AL Pic&0.548618(6)&34.13(6)&5.0&+&404&3287&34.07&dP14\\
NSV 1856&0.516080(2)&742(14)&8.4&&1018&3248&786.91&SF07\\
FR Psc&0.455680(4)&51.47(8)&5.8&&305&3240&51.31&dp14\\
NSV 3331=V0714 Pup&0.494182(2)&119.4(7)&7.2&&893&3272&116.96&SF07\\
BT Sco&0.548718(4)&39.11(8)&4.6&&508&3167&78&WS05\\
AR Ser&0.575213(8)&1325(60)&6.3&&463&3167&109&WS05\\
V5659 Sgr&0.379706(4)&45.58(8)&5.9&+&488&3134&45.7&WS05\\
ASAS J153830-6906.4&0.622472(6)&117.9(6)&5.9&+&660&3209&1702.13&SF07\\
EP Tuc&0.615003(6)&63.0(2)&4.0&&451&3274&63&WS05\\
BQ Vir&0.637027(4)&50.8(1)&4.4&&805&3160&-&K05\\
V0354 Vir&0.595045(8)&49.0(2)&4.0&&407&3168&59&W06\\
V0551 Vir&0.446854(8)&53.0(2)&4.6&+&289&2941&48&W06\\
V0574 Vir&0.474391(8)&25.71(4)&4.1&+&227&3099&26.3&B14\\
V0585 Vir&0.601611(4)&94.1(5)&4.7&+&574&3183&93.8&B14\\
V0586 Vir&0.682773(4)&134(1)&4.0&+&517&3174&132&B14\\
 \hline
	\end{tabular}\label{Tab:Known-detections}
	\tablefoot{Ref: K05 -- \citet{kovacs2005}; SF07 -- \citet{szczygiel2007}; B14 -- \citet{bramich2014}; W06 -- \citet{wils2006}; K11 -- \citet{khruslov2011}; WS05 -- \citet{wils2005}; dP13 -- \citet{deponthiere2013}; dP14 -- \citet{deponthiere2014}; BH15 -- \citet{bonnardeau2015}.}
	\end{table*}

\setlength\tabcolsep{4pt}
\renewcommand{\arraystretch}{0.9}
\begin{table*}
		\centering
		\caption{Stars of different variability type originally classified as RRab stars in VSX. Colour index is a rough value taken from \citet{zacharias2013}.}		
	\begin{tabular}{LLLMMLLL}\hline \hline 
ID&$P$ [d]&Type&Blend&\emph{B$-$V} [mag]&\emph{N}&\emph{TS} [d]&Remarks\\ \hline
NSVS 14248877&0.99842(4)&E:&+&1.2&344&2939 &\\
ASAS J134527-3616.8&4.77429(10)&CEP&&1.1&164&1008 &\\
ASAS J142204-3404.3&0.773789(6)&CEP&&1.0&141& 998 &\\
ASAS J020345-1729.1&4.089(12)&CEP&&1.1&455&3296 &\\
NSVS 15925156&1.06135(4)&CEP&&1.1&580&3293 &jump in amplitude at 2453500\\
ASAS J205934+1812.7&3.24610(4)&CEP&+&1.2&491&2422 &\\
ASAS J040615-3050.0&1.32465(4)&CEP&&0.9&593&3300 &\\
ASAS J144908-1150.8&119.907(2)&M&+&1.6&494&3174 &\\
CSS J153535.3-155042&0.121131(8)&DSCT&&0.9&499&2966 &amplitude only 0.1\,mag\\
ASAS J202746-2850.5&0.408462(4)&RRC/EW&&0.2& 701& 3250&additional period with 90\,d\\
ASAS J053552-0508.2&2.82362(4)&CEP&&1.2& 754& 3298&\\
ASAS J183333-6654.0&1.924273(4)&CEP&+&1.0&1066& 3162&\\
CSS J222816.3+192154&0.140082(14)&DSCT&&0.4& 292& 2366&amplitude only 0.1\,mag\\
ASAS J085816-3022.1&1.90949(4)&ELL&+&1.8& 598& 3291&amplitude only 0.1\,mag\\
NSVS 19229378&318(6)&LPV&+&0.7& 447& 3092&\\
AK Sct&189(4)&LPV&+&0.8& 348& 3119&\\
NSVS 19501853&319(2)&M&+&1.4&1476& 3191&\\
ASAS J135447-1640.8&0.37225(4)&EW&&0.8& 480& 3183&\\
ASAS J195215+1943.2&0.6169(6)&E:&+&0.5& 277& 2386&\\

 \hline
	\end{tabular}\label{Tab:OtherTypes}
	\end{table*}

\begin{figure*}[htbp!]
	\begin{center}
		\includegraphics[width=\hsize]{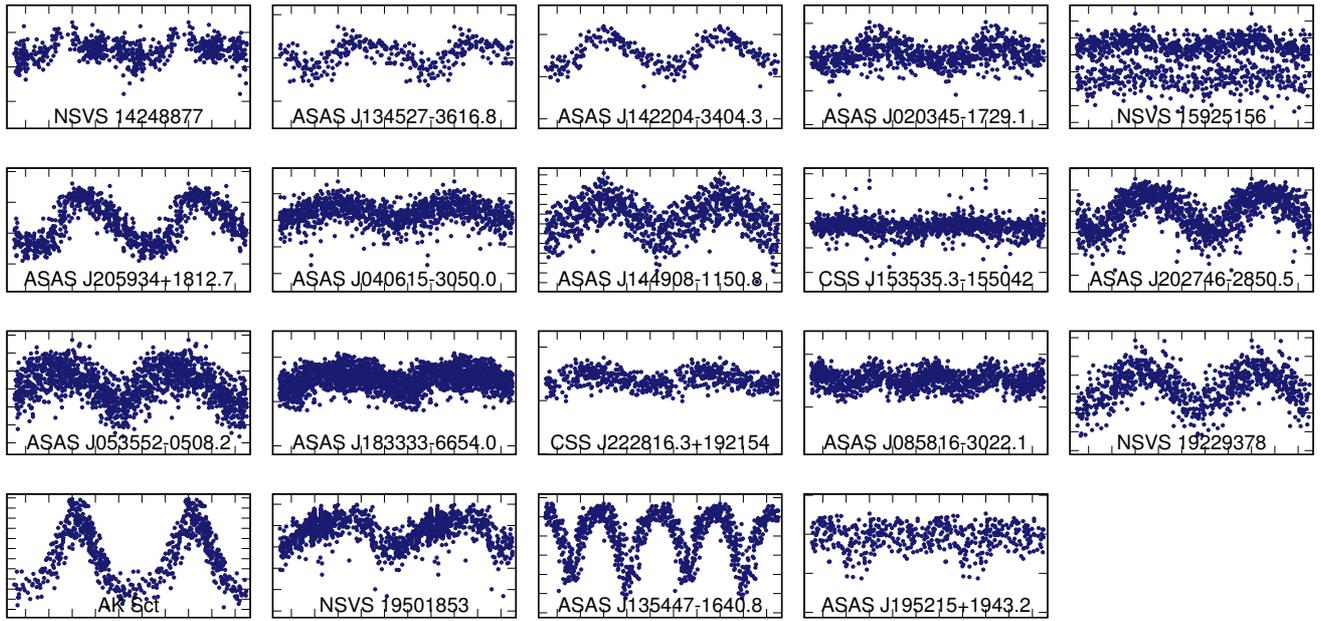}
		\caption{Stars misclassified as RRab stars in VSX identified in our sample. Data are phased with periods from Table \ref{Tab:OtherTypes}. The scales of axes are the same as in Fig. \ref{Fig:Mosaic}.}\label{Fig:Others}	
	\end{center}
\end{figure*}


\begin{thebibliography}{}
\bibitem[Alcock et al.(2000)]{alcock2000} Alcock, C., Allsman, R., Alves, D.~R., et al.\ 2000, \apj, 542, 257 
\bibitem[Alcock et al.(2003)]{alcock2003} Alcock, C., Alves, D.~R., Becker, A., et al.\ 2003, ApJ, 598, 597
\bibitem[Benk{\H o}, Szab{\'o}\,\&\,Papar{\'o}(2011)]{benko2011} Benk{\H o}, J.~M., Szab{\'o}, R., \& Papar{\'o}, M.\ 2011, MNRAS, 417, 974 
\bibitem[Benk{\H o} et al.(2014)]{benko2014} Benk{\H o}, J.~M., Plachy, E., Szab{\'o}, R., Moln{\'a}r, L., \& Koll{\'a}th, Z.\ 2014, \apjs, 213, 31 
\bibitem[Benk{\H o} \& Szab{\'o}(2015)]{benko2015} Benk{\H o}, J.~M., \& Szab{\'o}, R.\ 2015, European Physical Journal Web of Conferences, 101, 06008 
\bibitem[Bla{\v z}ko(1907)]{blazhko1907} Bla{\v z}ko, S.\ 1907, Astronomische Nachrichten, 175, 325
\bibitem[Bonnardeau \& Hambsch(2015)]{bonnardeau2015} Bonnardeau, M., \& Hambsch, F.-J.\ 2015, Information Bulletin on Variable Stars, 6132, 1 
\bibitem[Bramich et al.(2014)]{bramich2014} Bramich, D.~M., Alsubai, K.~A., Arellano Ferro, A., et al.\ 2014, Information Bulletin on Variable Stars, 6106, 1 
\bibitem[Chadid et al.(2010)]{chadid2010} Chadid, M., Benk{\H o}, J.~M., Szab{\'o}, R., et al.\ 2010, A\&A, 510, A39 
\bibitem[Collinge et al.(2006)]{collinge2006} Collinge, M.~J., Sumi, T., \& Fabrycky, D.\ 2006, \apj, 651, 197
\bibitem[de Ponthi{\`e}re et al.(2013)]{deponthiere2013} de Ponthi{\`e}re, P., Hambsch, F.-J., Krajci, T., \& Menzies, K., Wils, P.\ 2013, Journal of the American Association of Variable Star Observers (JAAVSO), 41, 58 
\bibitem[de Ponthi{\`e}re et al.(2014)]{deponthiere2014} de Ponthi{\`e}re, P., Hambsch, F.-J., Menzies, K., \& Sabo, R.\ 2014, Journal of the American Association of Variable Star Observers (JAAVSO), 42, 298
\bibitem[Detre \& Szeidl(1973)]{detre1973} Detre, L., \& Szeidl, B.\ 1973, Information Bulletin on Variable Stars, 764, 1
\bibitem[Goranskij, Clement\,\&\,Thompson(2010)]{goranskij2010} Goranskij, V., Clement, C.~M., \& Thompson, M.\ 2010, in Variable Stars, the Galactic Halo and Galaxy Formation, ed. C. Sterken, N. Samus, \& L. Szabados (Moscow: Sternberg Astronomical Institute of Moscow Univ.), 115
\bibitem[Guggenberger et al.(2012)]{guggenberger2012} Guggenberger, E., Kolenberg, K., Nemec, J.~M., et al.\ 2012, \mnras, 424, 649
\bibitem[Hajdu et al.(2015)]{hajdu2015} Hajdu, G., Catelan, M., Jurcsik, J., et al.\ 2015, \mnras, 449, L113
\bibitem[Jurcsik, S\'{o}dor\,\&\,V\'{a}radi(2005)]{jurcsik2005a} Jurcsik, J., S\'{o}dor, \'{A}, \& V\'{a}radi, M.\ 2005a, Information Bulletin on Variable Stars, 5666, 1 
\bibitem[Jurcsik et al.(2005)]{jurcsik2005b} Jurcsik, J., Szeidl, B., Nagy, A., \& S\'{o}dor, \'{A}\ 2005b, \actaa, 55, 303 
\bibitem[Jurcsik et al.(2009)]{jurcsik2009} Jurcsik, J., S{\'o}dor, {\'A}., Szeidl, B., et al.\ 2009, MNRAS, 400, 1006
\bibitem[Jurcsik et al.(2011)]{jurcsik2011} Jurcsik, J., Szeidl, B., Clement, C., Hurta, Z., \& Lovas, M.\ 2011, \mnras, 411, 1763
\bibitem[Jurcsik et al.(2012)]{jurcsik2012} Jurcsik, J., Hajdu, G., Szeidl, B., et al.\ 2012, \mnras, 419, 2173
\bibitem[Jurcsik et al.(2014)]{jurcsik2014} Jurcsik, J., Smitola, P., Hajdu, G., \& Nuspl, J.\ 2014, ApJL, 797, L3 
\bibitem[Jurcsik \& Smitola(2016)]{jurcsik2016} Jurcsik, J., \& Smitola, P.\ 2016, Commmunications of the Konkoly Observatory Hungary, 105, 167 
\bibitem[Khruslov(2011)]{khruslov2011} Khruslov, A.~V.\ 2011, Peremennye Zvezdy Prilozhenie, 11,
\bibitem[Koll{\'a}th, Moln\'{a}r\,\&\,Szab\'{o}(2011)]{kollath2011} Koll{\'a}th, Z., Moln{\'a}r, L., \& Szab{\'o}, R.\ 2011, MNRAS, 414, 1111 
\bibitem[Koci{\' a}n et al.(2012)]{kocian2012} Koci{\' a}n, R., Li{\v s}ka, J., Skarka, M., et al.\ 2012, Open European Journal on Variable Stars, 154, 1 
\bibitem[Kov{\'a}cs(2005)]{kovacs2005} Kov{\'a}cs, G.\ 2005, A\&A, 438, 227
\bibitem[Kov\'{a}cs(2015)]{kovacs2015} Kov\'{a}cs, G.\ 2015, arXiv:1512.05722
\bibitem[LaCluyz{\'e} et al.(2004)]{lacluyze2004} LaCluyz{\'e}, A., Smith, H.~A., Gill, E.-M., et al.\ 2004, \aj, 127, 1653
\bibitem[Le Borgne et al.(2007)]{leborgne2007} Le Borgne, J.~F., Paschke, A., Vandenbroere, J., et al.\ 2007, \aap, 476, 307
\bibitem[Le Borgne et al.(2014)]{leborgne2014} Le Borgne, J.~F., Poretti, E., Klotz, A., et al.\ 2014, \mnras, 441, 1435 
\bibitem[Lenz \& Breger(2005)]{lenz2005} Lenz, P., \& Breger, M.\ 2005, Communications in Asteroseismology, 146, 53 
\bibitem[Li{\v s}ka et al.(2015)]{liska2015} Li{\v s}ka, J., Skarka, M., Auer, R.~F., Prudil, Z., 
\& Jur{\'a}{\v n}ov{\'a}, A.\ 2015, Open European Journal on Variable Stars, 170, 1
\bibitem[Limpert, Stahel\,\&\,Abbt(2001)]{limpert2001} Limpert, E., Stahel, W.\,A., Abbt, M.\ 2001, BioScience, 51, 5
\bibitem[Moskalik \& Poretti(2003)]{moskalik2003} Moskalik, P., \& Poretti, E.\ 2003, \aap, 398, 213
\bibitem[Nagy \& Kov{\'a}cs(2006)]{nagy2006} Nagy, A., \& Kov{\'a}cs, G.\ 2006, \aap, 454, 257
\bibitem[Pigulski(2014)]{pigulski2014} Pigulski, A.\ 2014, Precision Asteroseismology, 301, 31 
\bibitem[Plachy et al.(2014)]{plachy2014} Plachy, E., Benk{\H o}, J.~M., Koll{\'a}th, Z., Moln{\'a}r, L., \& Szab{\'o}, R.\ 2014, \mnras, 445, 2810
\bibitem[Pojma{\'n}ski(1997)]{pojmanski1997} Pojma{\'n}ski, G.\ 1997, AcA, 47, 467
\bibitem[Pojma{\'n}ski(2001)]{pojmanski2001} Pojma{\'n}ski, G.\ 2001, IAU Colloq.~183: Small Telescope Astronomy on Global Scales, 246, 53 
\bibitem[Pojma{\'n}ski(2002)]{pojmanski2002} Pojma{\'n}ski, G.\ 2002, AcA, 52, 397
\bibitem[Pollacco et al.(2006)]{pollacco2006} Pollacco, D.~L., Skillen, I., Collier Cameron, A., et al.\ 2006, PASP, 118, 1407 
\bibitem[Poretti et al.(2016)]{poretti2016} Poretti, E., Le Borgne, J.-F., Klotz, A., Audejean, M., \& Hirosawa, K.\ 2016, Commmunications of the Konkoly Observatory Hungary, 105, 73
\bibitem[Skarka(2013)]{skarka2013} Skarka, M.\ 2013, A\&A, 549, A101 
\bibitem[Skarka(2014a)]{skarka2014a} Skarka, M.\ 2014a, A\&A, 562, A90
\bibitem[Skarka(2014b)]{skarka2014b} Skarka, M.\ 2014b, MNRAS, 445, 1584
\bibitem[Smolec(2016)]{smolec2016} Smolec, R.\ 2016, arXiv:1603.01252
\bibitem[S{\'o}dor et al.(2007)]{sodor2007} S{\'o}dor, {\'A}., Szeidl, B., \& Jurcsik, J.\ 2007, \aap, 469, 1033 
\bibitem[S{\'o}dor et al.(2012)]{sodor2012} S{\'o}dor, {\'A}., Hajdu, G., Jurcsik, J., et al.\ 2012, \mnras, 427, 1517
\bibitem[Soszy{\'n}ski et al.(2011)]{soszynski2011} Soszy{\'n}ski, I., Dziembowski, W.~A., Udalski, A., et al.\ 2011, AcA, 61, 1 
\bibitem[Szab{\'o}(2014)]{szabo2014a} Szab{\'o}, R.\ 2014, IAU Symposium, 301, 241
\bibitem[Szab{\'o} et al.(2014)]{szabo2014b} Szab{\'o}, R., Benk{\H o}, J.~M., Papar{\'o}, M., et al.\ 2014, \aap, 570, A100
\bibitem[Szab{\'o} et al.(2010)]{szabo2010} Szab{\'o}, R., Koll{\'a}th, Z., Moln{\'a}r, L., et al.\ 2010, MNRAS, 409, 1244
\bibitem[Szczygie{\l}\,\&\,Fabrycky(2007)]{szczygiel2007} Szczygie{\l}, D.~M., \& Fabrycky, D.~C.\ 2007, MNRAS, 377, 1263
\bibitem[Szeidl(1976)]{szeidl1976} Szeidl, B.\ 1976, IAU Colloq.~29: Multiple Periodic Variable Stars, 60, 133
\bibitem[Szeidl et al.(2011)]{szeidl2011} Szeidl, B., Hurta, Zs., Jurcsik, J., Clement, C., \& Lovas, M.\ 2011, \mnras, 411, 1744
\bibitem[Szeidl et al.(2012)]{szeidl2012} Szeidl, B., Jurcsik, J., S{\'o}dor, {\'A}., Hajdu, G., \& Smitola, P.\ 2012, \mnras, 424, 3094 
\bibitem[Walker \& Nemec(1996)]{nemec1996} Walker, A.~R., \& Nemec, J.~M.\ 1996, \aj, 112, 2026
\bibitem[Watson, Henden\,\&\,Price(2006)]{watson2006} Watson, C.~L., Henden, A.~A., \& Price, A.\ 2006, \\ 25th Annual Symposium, The Society for Astronomical Sciences, 47W
\bibitem[Wils \& Sodor(2005)]{wils2005} Wils, P., \& S\'odor, \'A.\ 2005, Information Bulletin on Variable Stars, 5655, 1 
\bibitem[Wils et al.(2006)]{wils2006} Wils, P., Lloyd, C., \& Bernhard, K.\ 2006, \mnras, 368, 1757
\bibitem[Zacharias et al.(2013)]{zacharias2013} Zacharias, N., Finch, C.~T., Girard, T.~M., et al.\ 2013, \aj, 145, 44
\end{thebibliography}
\end{document}